\documentclass[sigconf]{acmart}
\AtBeginDocument{%
  }

\setcopyright{acmlicensed}
\copyrightyear{2027}
\acmYear{2027}
\acmDOI{XXXXXXX.XXXXXXX}
\acmConference[SIGMOD '27]{Make sure to enter the correct
  conference title from your rights confirmation email}{June 13-19, 
  2027}{Huntington Beach, CA, USA}

\acmISBN{978-1-4503-XXXX-X/2018/06}

\newcommand{\ourSystemName}{HakiCC}
\newcommand{\ourSystem}{\emph{\ourSystemName}}

\newcommand{\cat}[1]{\smallskip\noindent\textbf{#1.}}

\newcommand{\circled}[1]{\raisebox{.5pt}{\textcircled{\raisebox{-.9pt}{#1}}}}

\usepackage{subcaption}
\usepackage{mathtools}
\usepackage{algorithm}
\usepackage{algpseudocode}
\usepackage{enumitem}
\usepackage{xcolor}
\usepackage{listings}
\usepackage{caption}
\usepackage{multirow}

\usepackage{afterpage} %

\lstdefinelanguage{CCPseudo}{
    morekeywords=[1]{Class,extends,State},
    morekeywords=[2]{begin,read,write,commit,rollback},
    morekeywords=[3]{Transaction,Value,CommitResult},
    morekeywords=[4]{return,success,abort},
    morekeywords=[5]{Override},
    sensitive=true,
    morecomment=[l]{//},
}
\author{Farzad Habibi}
\affiliation{%
  \institution{University of California, Irvine}
}
\email{habibif@uci.edu}

\author{Juncheng Fang}
\affiliation{%
  \institution{University of California, Irvine}
}
\email{junchf1@uci.edu}

\author{Faisal Nawab}
\affiliation{%
  \institution{University of California, Irvine}
}
\email{nawabf@uci.edu}

\makeatletter
\def\@ACM@checkaffil{%
    \if@ACM@instpresent\else
    \ClassWarningNoLine{\@classname}{No institution present for an affiliation}%
    \fi
    \if@ACM@citypresent\else
    \ClassWarningNoLine{\@classname}{No city present for an affiliation}%
    \fi
    \if@ACM@countrypresent\else
        \ClassWarningNoLine{\@classname}{No country present for an affiliation}%
    \fi
}
\makeatother
\begin{document}
\title{\ourSystemName: LLM-Driven Multi-Agent Design and Optimization of Concurrency Control Protocols}

\begin{abstract}
Large language models (LLMs) have recently been applied in systems research as a tool to reduce human-intensive engineering effort through cost-efficient automation.
Decades of research have produced a rich landscape of concurrency control (CC) protocols, each encoding distinct trade-offs in correctness, throughput, and abort behavior.
However, most applications in practice default to 2PL or OCC, because selecting and adapting a protocol to a specific application requires expert knowledge that is rarely available to application designers.
This is a wasted opportunity, as an application-specific CC protocol can yield significant performance advantages over a generic baseline, but designing one requires deep expertise in CC protocol design.

In this paper, we propose \ourSystem, an LLM-driven multi-agent pipeline that automatically designs, verifies, and optimizes concurrency control protocols tailored to a given target application.
\ourSystem\ provides a two-stage pipeline. In Stage~1, a multi-agent system takes a workload description as input and generates an application-specific CC protocol implementation, which is iteratively repaired and verified for conflict-serializability. In Stage~2, the verified protocol is further optimized for that application through an LLM-driven evolutionary loop targeting correctness and throughput.
We evaluate \ourSystem\ on TPC-C and AuctionMark as target workloads, producing and reporting ten application-specific CC protocols. All ten are conflict-serializable after Stage~1; Stage~2 improves throughput for nine out of the ten protocols, with average gains of +50.6\% for TPC-C protocols and +92.2\% for AuctionMark protocols.

\end{abstract}


\settopmatter{printfolios=true}

\begin{abstract}
Large language models (LLMs) have recently been applied in systems research as a tool to reduce human-intensive engineering effort through cost-efficient automation.
Decades of research have produced a rich landscape of concurrency control (CC) protocols, each encoding distinct trade-offs in correctness, throughput, and abort behavior.
However, most applications in practice default to 2PL or OCC, because selecting and adapting a protocol to a specific application requires expert knowledge that is rarely available to application designers.
This is a wasted opportunity, as an application-specific CC protocol can yield significant performance advantages over a generic baseline, but designing one requires deep expertise in CC protocol design.

In this paper, we propose \ourSystem, an LLM-driven multi-agent pipeline that automatically designs, verifies, and optimizes concurrency control protocols tailored to a given target application.
\ourSystem\ provides a two-stage pipeline. In Stage~1, a multi-agent system takes a workload description as input and generates an application-specific CC protocol implementation, which is iteratively repaired and verified for conflict-serializability. In Stage~2, the verified protocol is further optimized for that application through an LLM-driven evolutionary loop targeting correctness and throughput.
We evaluate \ourSystem\ on TPC-C and AuctionMark as target workloads, producing and reporting ten application-specific CC protocols. All ten are conflict-serializable after Stage~1; Stage~2 improves throughput for nine out of the ten protocols, with average gains of +50.6\% for TPC-C protocols and +92.2\% for AuctionMark protocols.

\end{abstract}

\maketitle

\section{Introduction}
\label{sec:intro}

Concurrency control (CC) is a foundational component of Online Transaction Processing (OLTP) systems, defining how concurrent transactions access shared data while preserving correctness and maximizing performance~\cite{bernstein1987concurrency}.
Decades of research have produced a rich landscape of protocol classes, each with hundreds of variants, including Two-Phase Locking (2PL)~\cite{bernstein1987concurrency}, Optimistic Concurrency Control (OCC)~\cite{kung1981occ}, Multi-Version Concurrency Control (MVCC)~\cite{bernstein1983multiversion, levandoski2015multi}, and Serializable Snapshot Isolation (SSI)~\cite{cahill2008ssi}, each encoding a distinct trade-off between blocking behavior, validation overhead, abort rate, and correctness guarantees~\cite{yu2014ccEvaluation}.
Despite this variety, designing a new application-specific CC protocol remains expert-driven and labor-intensive, since engineers must reason simultaneously about correctness under concurrency, performance under target workloads, and implementation correctness, all without automated assistance. Protocol optimizations such as commutativity-aware conflict resolution~\cite{oneil1986escrow}, static transaction analysis~\cite{wang2016ics3, habibi2026brook, fang2026immortalchopper}, and deterministic pre-ordering~\cite{thomson2012calvin, faleiro2014bohm} all require deep knowledge of both the application's transaction logic and the underlying CC mechanism, making them difficult to scale without domain expertise.

LLM-driven agents have recently demonstrated the ability to reason about workloads and protocols, generate executable implementations, and iteratively repair them under feedback~\cite{chen2021codex, shinn2023reflexion, yang2024sweagent}.
LLM-driven evolutionary frameworks extend this capability by maintaining a population of candidate programs, using the LLM to propose mutations, and selecting survivors via a benchmark harness, enabling automated exploration of complex design spaces that are too large for exhaustive search~\cite{novikov2025alphaevolve, openevolve, assumpccao2025codeevolve, romera2024funSearch}.
These developments raise a question. \textit{Can an LLM-driven pipeline design an application-specific concurrency control protocol from scratch, verify its correctness, and optimize it for performance without human feedback in the loop?}

We answer this question with \ourSystem, a two-stage multi-agent pipeline for automated application-specific CC protocol design and optimization.
In \textbf{Stage~1}, a multi-agent system takes a target workload as input, designs a protocol tailored to it, generates an executable implementation, and iteratively repairs the implementation until it passes correctness verification.
In \textbf{Stage~2}, an LLM-driven evolutionary optimizer refines the verified protocol through repeated cycles of mutation and benchmark evaluation, resulting in an optimized protocol.

We instantiate \ourSystem\ on two standard OLTP benchmarks---TPC-C~\cite{tpc-c} and AuctionMark~\cite{DifallahPCC13}---producing and reporting five protocols per workload.
All ten generated protocols are conflict-serializable.
Structural analysis across 150 binary design criteria shows that the generated protocols span multiple design families: the five TPC-C protocols generated by \ourSystem\ (SALVO, LP-OCC, PESC, DCOV, SADL) mostly cluster around the OCC/Silo~\cite{tu2013speedy, kung1981occ} family, with SADL being the exception by combining deterministic pre-ordering with commutative escrow locking. The five AuctionMark protocols generated by \ourSystem\ (TREAD, WGC, HSL, PVL, SAPL) include an MVCC-based design (TREAD), a deterministic partitioned-scheduling design (WGC), two OCC-family designs (HSL, PVL), and a 2PL-OCC-Escrow hybrid (SAPL).
Stage~2 optimization improves throughput for most of the ten protocols, with average gains of +50.6\% for TPC-C protocols and +92.2\% for AuctionMark protocols, while maintaining correctness.

This paper makes the following contributions.
\begin{itemize}[leftmargin=*, itemsep=2pt, topsep=2pt]
  \item We introduce \ourSystem, the first end-to-end two-stage LLM-driven pipeline that designs, implements, verifies, and optimizes application-specific concurrency control protocols.
  \item We demonstrate that the pipeline produces correct protocols. All ten generated implementations pass our conflict-serializability verifier, and Stage~2 optimization improves performance without sacrificing correctness in nine out of the ten cases.
  \item We show that application-conditioned design induces structural diversity across generated protocols, spanning pessimistic, optimistic, versioned, deterministic, and semantic design paradigms, as confirmed by pairwise Jaccard similarity analysis over 150 design criteria.
  \item We describe ten CC protocols generated by \ourSystem\ and catalog the concrete improvements discovered by Stage~2 evolution.
\end{itemize}

Section~\ref{sec:Background} provides background, Section~\ref{sec:design} presents the \ourSystem\ design, Section~\ref{sec:experiments} reports the evaluation, Section~\ref{sec:protocols} describes the ten generated protocols, Section~\ref{sec:Related_work} discusses related work, and Section~\ref{sec:conclusion} concludes the paper.

\section{Background and Motivation}
\label{sec:Background}

\subsection{Database Transactions}

A database transaction is a sequence of operations executed as a single unit of work, ensuring the ACID properties (atomicity, consistency, isolation, and durability). A concurrent schedule is serializable if it is equivalent to some serial execution of its transactions; conflict-serializability is a stricter form that requires preserving the relative order of all conflicting operations~\cite{bernstein1987concurrency}.

\cat{Concurrency Control Protocols}
Concurrency control protocols govern the interleaving of concurrent transactions, ensuring conflict-serializability~\cite{bernstein1987concurrency}.
Prior work has produced a diverse protocol landscape. Two-Phase Locking (2PL) prevents conflicts by acquiring locks before releasing them; Optimistic Concurrency Control (OCC)~\cite{kung1981occ} defers conflict detection to commit time; Multi-Version Concurrency Control (MVCC)~\cite{bernstein1983multiversion} maintains multiple data versions to allow readers to proceed without blocking writers; and Serializable Snapshot Isolation (SSI)~\cite{cahill2008ssi} extends snapshot isolation to full serializability via anti-dependency tracking.
More recent hybrid designs combine primitives from multiple families, for example, pairing optimistic validation with commutative-operation escrow~\cite{oneil1986escrow} to reduce abort rates on counter-heavy workloads.

Despite this variety, designing a new CC protocol remains a fundamentally human-driven process, requiring simultaneous decisions about conflict detection timing, version management, abort policy, and application-specific optimizations that interact in non-obvious ways.

\cat{Correctness Verification}
A candidate CC protocol must be evaluated not only for performance but also for correctness under concurrent execution.
For conflict-serializability, correctness can be checked by constructing a \textit{precedence graph} $G_H = (\mathcal{T}, E)$ over an execution trace, where each node is a committed transaction and each edge $(T_i, T_j)$ represents a conflicting operation pair in execution order~\cite{bernstein1987concurrency, papadimitriou1979serializability}.
An execution is conflict-serializable if and only if $G_H$ is acyclic; a detected cycle constitutes a serializability violation.
\ourSystem\ uses precedence-graph cycle detection as its correctness verification mechanism.
This check is empirical and schedule-dependent, so passing it certifies serializability on the observed interleavings rather than all of them; we mitigate this with a heavily concurrent verification suite designed to surface unserializable behavior.

\subsection{LLM-Driven Agentic Optimization}
\cat{Agentic Workflows}
Machine learning has long been used in systems research to guide design choices, tune configuration parameters, and model workload behavior~\cite{jay2018ml, van2017ml, yang2022ml, kraska2018ml, marcus2019ml, wang2021polyjuice, pan2026modeling}.
More recently, LLM-based coding agents have demonstrated an ability to rapidly prototype implementations, propose algorithmic variations, and repair code under structured feedback, substantially reducing the manual engineering effort required for complex software tasks~\cite{chen2021codex, shinn2023reflexion, yang2024sweagent, langchain2022, langgraph2024, claudecode2025}.
These capabilities extend naturally to multi-step agentic workflows, where LLM calls are composed into pipelines that handle tasks requiring iterative reasoning, code generation, and feedback-driven repair.

Systems algorithms in general, and CC protocols in particular, are well-suited for automated design because their quality can be evaluated against objective, well-defined metrics such as throughput for performance and conflict-serializability for correctness~\cite{cheng2025barbarians, cheng2025let, cheng2026aiDatabases}. This closed evaluation loop eliminates the need for human judgment when scoring a candidate, making the labor-intensive process of proposing, implementing, and refining protocols a natural fit for LLM-driven agentic workflows.

\cat{Evolutionary Frameworks}
Beyond code generation, recent work has begun to automate portions of the research process itself, iteratively proposing hypotheses, running experiments, and refining designs through feedback-driven agentic loops.

A particularly powerful instantiation of this idea is the LLM-driven evolutionary framework~\cite{romera2024funSearch,novikov2025alphaevolve,assumpccao2025codeevolve}, in which an LLM maintains a population of candidate solutions (code, designs, or algorithms) and iteratively improves them through an evolutionary loop.
In each iteration, the LLM generates and mutates candidates; an evaluator scores them via tests, benchmarks, or formal verification; and the highest-scoring candidates seed the next round.
This loop combines the generative capabilities of LLMs with the systematic coverage of evolutionary search, enabling exploration of design spaces that are too large for exhaustive enumeration and too structured for purely random search.

\ourSystem\ builds on this paradigm, applying it to CC protocol optimization rather than initial design. The CC protocols' design space is structured by design principles; however, it is too large for exhaustive search, making LLM-driven evolutionary refinement a better fit for improving existing protocol implementations against performance benchmarks. Rather than designing protocols from scratch, this evolutionary optimization takes a protocol produced by \ourSystem's agentic design workflow as its starting point and evolves that implementation for performance.

\begin{figure*}[t]
    \centering
    \includegraphics[width=\textwidth]{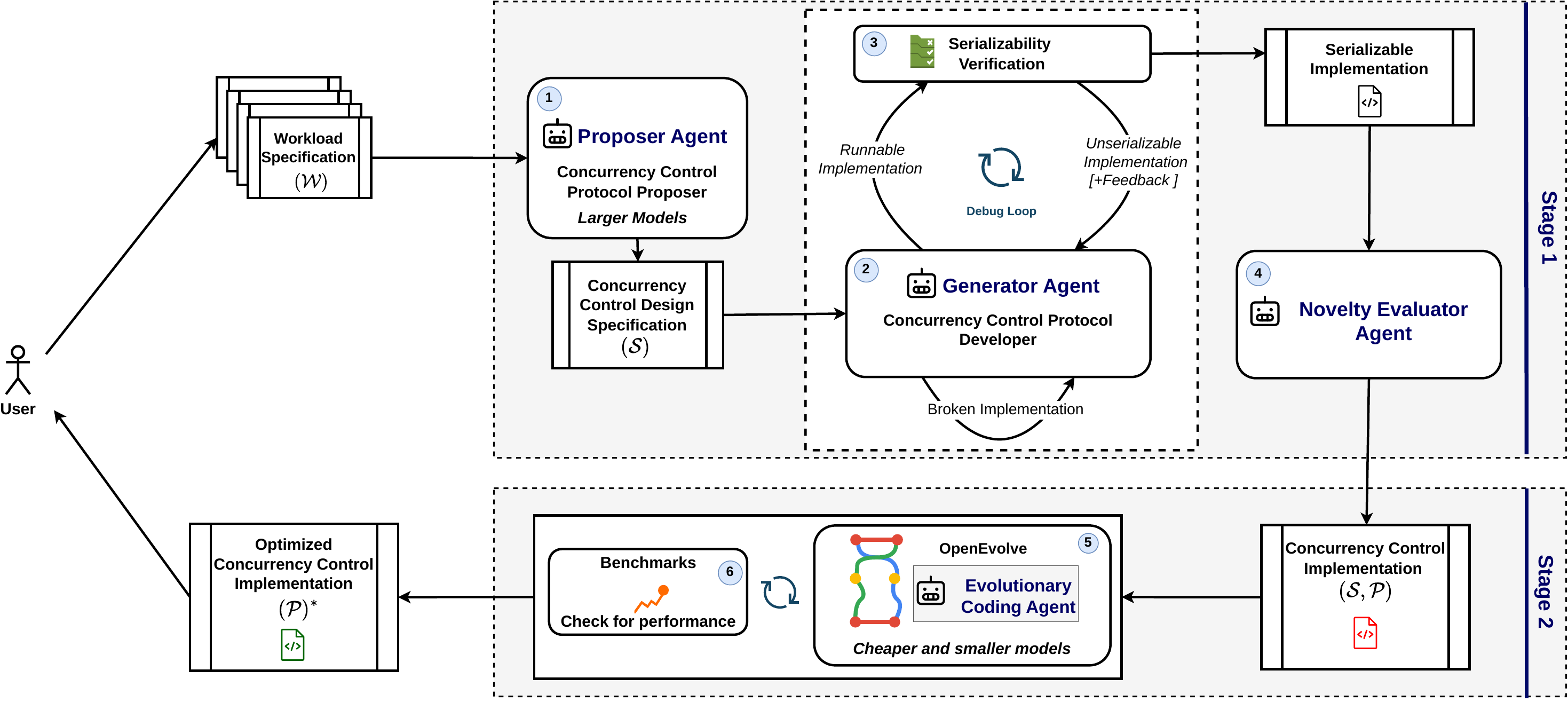}
    \caption{Overview of the \ourSystem\ two-stage pipeline. \textbf{Stage~1} (top): the \textit{Proposer Agent}~(\circled{1}) generates a design specification $\mathcal{S}$ from workload input $\mathcal{W}$; the \textit{Generator Agent}~(\circled{2}) translates $\mathcal{S}$ into an executable implementation; the \textit{Serializability Verifier}~(\circled{3}) checks correctness and returns failure traces to the Generator via a debug loop; and the \textit{Novelty Evaluator Agent}~(\circled{4}) scores and packages the verified protocol $\mathcal{P}$. \textbf{Stage~2} (bottom): OpenEvolve~(\circled{5}) iteratively optimizes $\mathcal{P}$ against performance benchmarks~(\circled{6}), yielding the final optimized protocol $\mathcal{P}^*$ returned to the user.}
    \label{fig:pipeline}
\end{figure*}

\section{\ourSystem\ Design}
\label{sec:design}
We present \ourSystem, an LLM-driven multi-agent pipeline that creates, implements, and iteratively improves concurrency control protocols with minimal human intervention.
Figure~\ref{fig:pipeline} shows the \ourSystem\ pipeline, which operates in two stages: the first designs and implements a correct concurrency control protocol based on application specification; the second optimizes it for performance using an agentic evolutionary framework.
While we focus on concurrency control as a concrete and well-scoped target, the pipeline can be adapted to other distributed data management protocols. With minimal modifications to workloads and correctness checks, the same framework supports AI-assisted protocol discovery and refinement more broadly.

\subsection{Stage 1: Proposal Pipeline}
Stage~1 designs and implements a correct concurrency control protocol tailored to a given workload specification---a set of transaction types and a target database schema.
The stage takes as input the workload specification and produces as output a verified, conflict-serializable protocol ready to serve as an input for the evolutionary optimization in Stage~2.
To maximize design quality, Stage~1 employs larger LLMs and typically completes within minutes.

\subsubsection{Workflow}
Stage~1 takes as input an application specification $\mathcal{W}$---a set of transaction types and stored procedures---and produces as output a verified, executable CC protocol implementation $\mathcal{P}$ that is conflict-serializable under the verification suite.
We model the pipeline as a stateful directed graph $G = (V, E)$~\cite{langgraph2024}, where each node is either an LLM invocation or a program execution. The pipeline consists of four components: a \textit{Proposer} agent (labeled \circled{1} in Figure~\ref{fig:pipeline}) that generates a protocol specification, a \textit{Generator} agent (\circled{2}) that translates it into executable code, a \textit{Serializability Verifier} (\circled{3}) that checks correctness and routes failures back to the Generator agent, and a \textit{Novelty Evaluator} agent (\circled{4}) that scores the verified protocol.

\cat{Step 1} The \textit{Proposer} (\circled{1}) agent takes $\mathcal{W}$ (application-specific workload) as input and generates a protocol specification $\mathcal{S}$, which defines read/write visibility rules, commit logic, expected correctness guarantees, and required metadata.
The agent is prompted as \textit{an expert in database concurrency control research and design} and is given $\mathcal{W}$ alongside instructions to design a CC protocol optimized for the given application, in the required output format. The format mandates a protocol name, a core design idea, a performance rationale tied to the target transactions, a correctness justification, and a step-by-step execution description covering transaction begin, read/write operations, conflict handling, and commit/rollback logic. Requiring the Proposer to justify correctness pushes it toward designs it can defend as serializable, leading to more serializable designs. The Proposer is additionally given the protocols \ourSystem\ has already produced for the application and instructed to differ from them, steering generation toward unexplored design space rather than re-generating prior solutions.

\cat{Step 2} The \textit{Generator} (\circled{2}) agent takes $\mathcal{S}$ as input and generates an executable implementation $\mathcal{P}$.
The Generator is prompted as \textit{an expert software engineer specializing in database concurrency control protocols}, with instructions to implement the protocol specification $\mathcal{S}$ as a concrete subclass of the provided concurrency control interface in the codebase. The interface defines five abstract methods---\texttt{begin}, \texttt{read}, \texttt{write}, \texttt{commit}, and \texttt{rollback} (see Figure~\ref{lst:cc_template}, discussed in Section~\ref{sec:final_stage_1_output}).

If no syntactically valid implementation is extracted, the Generator re-invokes the LLM and iterates until one is produced. The implementation $\mathcal{P}$ is then passed to the next step, \textit{Serializability Verifier} \circled{3}. Failures are returned to the Generator for repair.

Separating proposal from implementation allows each component to use an LLM suited to its demands: the Proposer can invoke a frontier-class model to maximize design quality, while the Generator can use a model optimized for code synthesis and repair. This decoupling also allows the Proposer to reason about protocol logic without being constrained by implementation details, fostering more structurally diverse designs.

\begin{figure}[t]
\begin{lstlisting}[frame=single, tabsize=2,
    belowskip=-8pt, aboveskip=4pt]
{Protocol Summary Report ($\mathcal{S}$)}

Class <CC_NAME> extends ConcurrencyControl:
  State:
    // protocol-specific (locks, versions, timestamps, ...)

  @Override
  begin(database) $\to$ Transaction:
    // initialize transaction context

  @Override
  read(txn, key, database) $\to$ Value:
    // acquire read permission; return value

  @Override
  write(txn, key, value, database):
    // acquire write permission; buffer updates; apply updates

  @Override
  commit(txn, database) $\to$ CommitResult:
    // validate; apply buffered writes; release state
    return success | abort

  @Override
  rollback(txn, database):
    // undo writes; release state
\end{lstlisting}
\caption{Template for the Stage~1 output: a concrete subclass of the \texttt{ConcurrencyControl} abstract class. The LLM fills in the \textit{State} block and all method bodies according to the protocol specification $S$ produced by the Proposer.}
\label{lst:cc_template}
\end{figure}

\cat{Step 3} The \textit{Serializability Verifier} (\circled{3}) receives $\mathcal{P}$ and evaluates it against a verification suite of concurrent workloads (detailed in Section~\ref{sec:setup}), each executed by $N$ parallel worker threads for a duration of $\tau$.

Correctness is checked by constructing a precedence graph $G_H = (\mathcal{T}, E)$ over the operation trace, where $\mathcal{T}$ is the set of committed transactions and $(T_i, T_j) \in E$ if and only if $T_i$ and $T_j$ have a conflicting operation pair in execution order~\cite{bernstein1987concurrency}. An execution is conflict-serializable if and only if $G_H$ is acyclic; a detected cycle constitutes a serializability violation. If any workload produces a cycle or a runtime failure, a structured error report---including the trace and cycle description---is fed back to the Generator agent (\circled{2}), which retakes control and attempts a repair. This generate-and-verify loop continues until all workloads pass, at which point the verified $\mathcal{P}$ is forwarded to the next step.

\cat{Step 4} In the next step, the \textit{Novelty Evaluator} (\circled{4}) agent serves as the final quality check in Stage~1.
This agent scores the verified protocol $\mathcal{P}$ by prompting an LLM as \textit{an expert database researcher and judge} to evaluate it against both established CC designs and the protocols \ourSystem\ has already generated for the application, so that redundancy with either is penalized and the accepted protocols stay diverse.
The judge independently scores four dimensions and assigns one penalty, each on a $[0,1]$ scale, which are accumulated into the final novelty score $\mathcal{N}(\mathcal{P})$:
\begin{equation}
    \mathcal{N}(\mathcal{P}) = 0.40\,n_{\text{orig}} + 0.30\,n_{\text{dist}} + 0.15\,n_{\text{impl}} + 0.15\,n_{\text{arg}} - 0.15\,p_{\text{rebrand}}
    \label{eq:novelty}
\end{equation}
where $n_{\text{orig}}$ is conceptual originality, $n_{\text{dist}}$ is distance from known protocol families, $n_{\text{impl}}$ is implementation distinctiveness, $n_{\text{arg}}$ is originality of the correctness argument, and $p_{\text{rebrand}} \in [0,1]$ is a penalty assigned if the design merely renames or marginally tunes an existing scheme. The accumulated score $\mathcal{N}(\mathcal{P})$ informs termination and restart decisions (discussed in Section~\ref{sec:termination}).

\subsubsection{Final Stage~1 Output}
\label{sec:final_stage_1_output}
Figure~\ref{lst:cc_template} shows the template of the Stage~1 output in \ourSystem. The output is a complete, runnable concurrency control protocol ready to deploy in our transactional system for verification and benchmarking. It begins with the protocol summary specification $\mathcal{S}$ followed by the generated code as a concrete subclass of an abstract class. Enforcing a fixed \texttt{begin} / \texttt{read} /\texttt{write} / \texttt{commit} / \texttt{rollback} interface keeps the output coherent. It simplifies benchmarking and verification, reduces the tokens the LLM must generate, and accelerates the repair loop by confining the task to implementing a single class.

\begin{figure}
    \centering
    \includegraphics[width=0.9\linewidth]{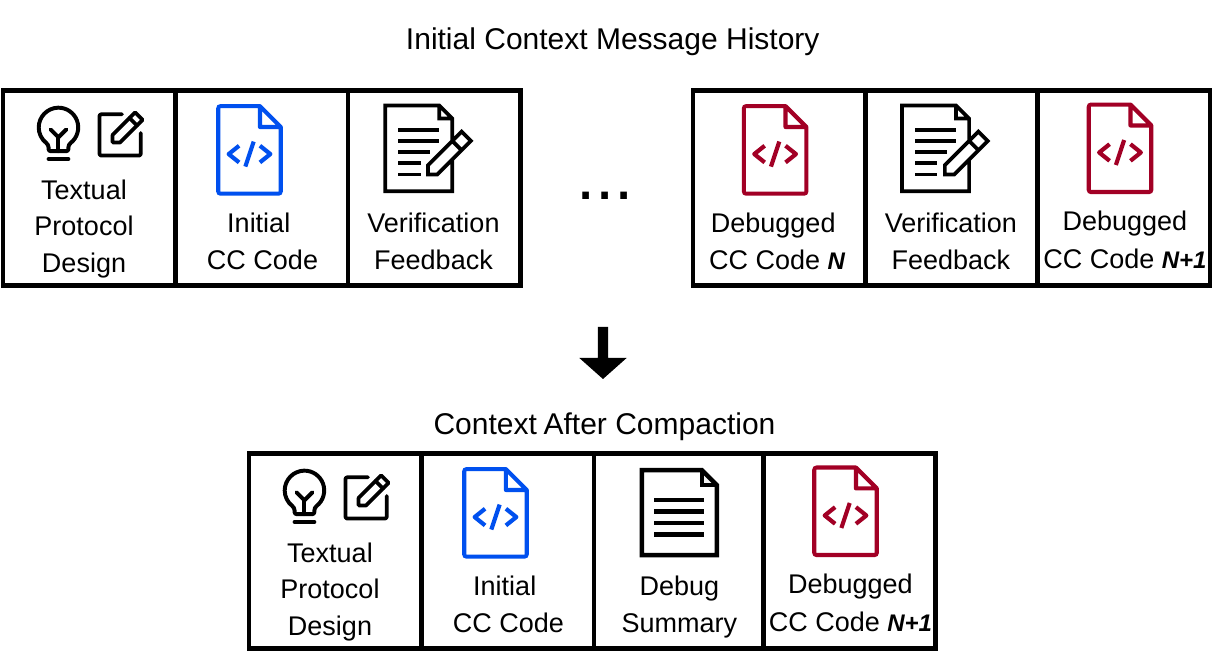}
    \caption{\ourSystem\ context compaction strategy.}
    \label{fig:contextHistory}
\end{figure}

\subsubsection{Shared Conversation History}
All nodes communicate through a shared conversation history $H$. Each node appends its output to $H$; every subsequent node receives the full accumulated context from all prior steps. This makes all prior outputs visible to every downstream component without explicit routing.
Sharing $H$ lets each agent consult the reasoning behind earlier outputs rather than only the artifact they produced, so the Generator agent can resolve ambiguities in $\mathcal{S}$ against the Proposer's stated intent and avoid retrying repairs that have already failed.

\subsubsection{LLM Context Management}
Each repair iteration appends error traces to $H$, causing the Generator's effective context to grow with the number of iterations. We bound this growth with a context compaction strategy (Figure~\ref{fig:contextHistory}): once $|H|$ exceeds a configurable token threshold, older debug entries are summarized and dropped, retaining only the original specification $S$ and the most recent error feedback. This keeps the Generator's context size complexity to be $O(1)$ in the number of repair iterations.

\subsubsection{Termination and Restart} 
\label{sec:termination}
The repair loop is bounded by a maximum iteration limit $k_{\max}$. If the \textit{Generator} agent fails to produce a passing implementation within $k_{\max}$ iterations, the pipeline discards $\mathcal{P}$ and restarts from the \textit{Proposer} agent with a fresh invocation.
A restart is also triggered if the novelty score $\mathcal{N}(\mathcal{P})$ falls below a configurable threshold $\theta_{\mathcal{N}}$, enforcing a minimum quality bar on the generated design before it advances to Stage~2.

\subsection{Stage 2: Agentic Optimization Pipeline}
Given the verified protocol $\mathcal{P}_0$ from Stage~1, Stage~2 solves the optimization problem:
\begin{equation}
    \mathcal{P}^* = \arg\max_{\mathcal{P} \in \Pi}\; \text{score}(\mathcal{P})
    \label{eq:opt}
\end{equation}
where $\Pi$ is the space of conforming CC implementations (i.e., subclasses of the CC interface that compile and run) and $\text{score}$ is defined in Equation~\ref{eq:score}. This score combines two objectives, the correctness of the protocol and its throughput, so that the search rewards implementations that are both serializable and fast. Rather than treating correctness as a hard constraint on $\Pi$, we keep it as a graded term of the score, so partially correct candidates are retained during the search and can still guide it toward a correct, high-performance design; as we detail below, the protocol \ourSystem\ ultimately returns is filtered to be fully conflict-serializable.
We implement this search using OpenEvolve~\cite{openevolve}, an open-source LLM-driven evolutionary framework. The optimizer iteratively proposes mutations of candidate implementations, evaluates them under controlled workloads, and selects high-performing variants to seed the next generation.

\cat{Evolutionary Search}
OpenEvolve agent (labeled \circled{5} in Figure~\ref{fig:pipeline}) optimizes the Stage~1 protocol through an iterative LLM-driven evolutionary loop. The search maintains a diverse population of candidate implementations using MAP-Elites (Multidimensional Archive of Phenotypic Elites)~\cite{mouret2015mapelites}, a quality-diversity algorithm that maps programs to behavioral cells and retains diverse high-performing variants rather than converging on a single local optimum. To prevent premature convergence, the population is split into multiple isolated islands that evolve independently, with periodic migration to propagate discoveries across the population. Within each island, selection balances exploitation of currently high-performing programs against exploration of novel variants.

Evolution is \emph{diff-based}, meaning the LLM proposes targeted, localized edits to the current best program rather than full rewrites. Restricting mutations to diffs preserves the core algorithmic structure established in Stage~1 and confines the search to the neighborhood of a known-correct protocol, directing optimization toward performance improvements rather than rediscovering correctness from scratch.

\cat{Evaluation Benchmark Suite}
Each candidate is evaluated against a benchmark suite of concurrent workloads (\circled{6}) that tests both correctness and performance. The correctness workloads check that the protocol remains conflict-serializable and makes progress under different scenarios. The performance workloads use standard OLTP benchmarks---TPC-C~\cite{tpc-c} and AuctionMark from OLTP-Bench~\cite{DifallahPCC13}---each run at multiple contention levels ranging from uniform random access to extreme hotspot scenarios, ensuring that optimized protocols generalize across typical and stress-test conditions. The correctness suite additionally includes three targeted synthetic workloads: a \textit{banking} workload (concurrent fund transfers), a \textit{counter} workload (read-modify-write increments on a shared counter), and a \textit{deadlock} workload (cyclic resource-acquisition patterns). The full benchmark configuration is detailed in Section~\ref{sec:experiments}.

\cat{Optimization Objective}
Each candidate is scored on two primary dimensions: \textbf{correctness}, which measures the fraction of workload scenarios in which the protocol is conflict-serializable, stall-free, and makes measurable progress over the duration of each run; and \textbf{throughput}, which captures the committed transaction rate across scenarios.
The primary optimization target is a \textbf{combined score}:
\begin{equation}
    \text{score} = 0.5 \cdot \text{correctness} + 0.5 \cdot \hat{T}
    \label{eq:score}
\end{equation}
where $\hat{T}$ is a weighted average of per-scenario throughputs, each normalized by a fixed throughput ceiling. This ceiling is the same for all workloads and set to the system's throughput capacity for a normal workload, so a protocol that reaches the maximum attainable throughput under a normal workload scores $\hat{T}=1$.
Scenarios involving the \emph{pre-known target application} (i.e., the application transaction mix the protocol is being optimized for) receive twice the weight in $\hat{T}$, biasing the search toward protocols that perform well on the intended use case rather than incidentally on other benchmarks.

This joint objective lets the search improve throughput while keeping correctness in view at every step, rather than optimizing performance in isolation and checking correctness only afterward. Because correctness is measured over a small suite of adversarial scenarios, it behaves as a near-binary signal in practice, as a single non-serializable scenario sharply lowers the score and an implementation bug typically breaks several scenarios at once. Starting from a correct Stage~1 protocol, any loss of correctness thus carries a penalty far larger than a plausible throughput gain, so the search is steered toward correct designs while still receiving a gradient from partially correct intermediates. At the end of the run, we filter the population by correctness and return the best-scoring protocol among those that pass all scenarios, so the final design is always conflict-serializable.

\section{Evaluation}
\label{sec:experiments}

In this section, we evaluate \ourSystem\ on two standard OLTP benchmarks, TPC-C and AuctionMark. We ran the full pipeline on each application and report the first ten protocols it produced, five per workload. For Stage~1, we examine the structural diversity of the generated protocols to assess whether application-conditioned design produces a meaningful spread across the CC design space. For Stage~2, we compare the performance of the generated protocols against each other and against  2PL and OCC baselines.

\begin{figure*}[t]
    \centering
    \includegraphics[width=\textwidth]{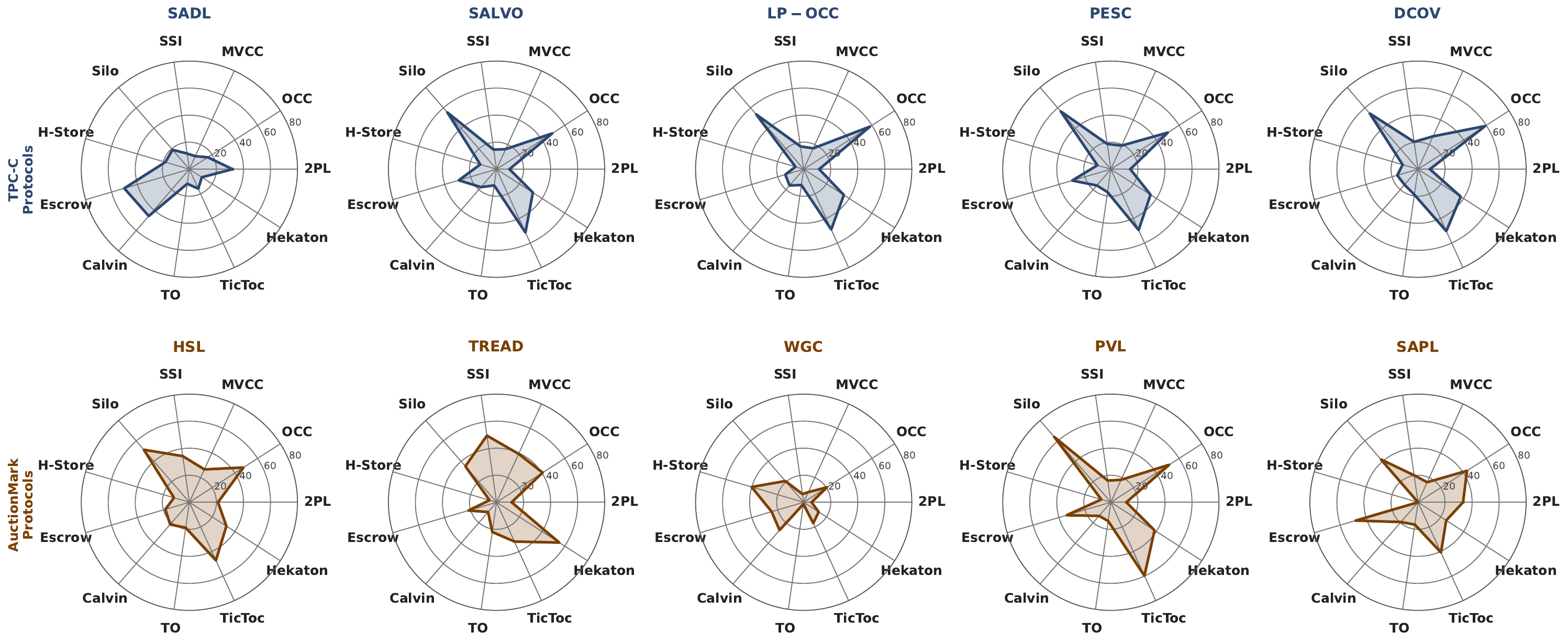}
    \caption{Design similarity profiles of the ten generated CC protocols against eleven reference CC families. Each radial axis reports the Jaccard similarity (\%) between a generated protocol's design feature vector and that of the reference family.}
    \label{fig:comparison}
\end{figure*}
\subsection{Experimental Setup}
\label{sec:setup}

We describe the evaluation configurations for each stage, the benchmark suite used for both correctness checking and performance evaluation, and the hardware platform.

\cat{Stage 1 Configuration}
All Stage~1 agents (\textit{Proposer},  \textit{Generator}, and \textit{Novelty Evaluator}) are backed by \textit{Gemini 2.5 Pro} at temperature 0.7 with a context window of up to 120K tokens.

The repair loop is capped at $k_{\max} = 50$ iterations; if the Generator agent fails to produce a passing implementation within this budget, the pipeline discards the current proposal and restarts from the Proposer. A restart is also triggered if the novelty score $\mathcal{N}(\mathcal{P})$ falls below $\theta_{\mathcal{N}}=0.25$ (see Section~\ref{sec:termination} for the full termination and restart policy).

\cat{Stage 2 Configuration}
The evolutionary optimization in Stage~2 is configured as follows.
We run OpenEvolve for 1000 iterations, evaluating candidates serially. 
The LLM backbone is \textit{Gemini 2.5 Flash} (temperature 0.7, max 120K tokens per call), and evolution is strictly \emph{diff-based}: only localized edits are proposed; full rewrites are disabled.

The population is structured as 3 independent islands, each maintaining an archive of 100 MAP-Elites cells within a global population cap of 800 programs.
Islands migrate periodically: every 50 generations, 15\% of each island's programs are shared with neighboring islands.
The \emph{exploration ratio} (30\%) and \emph{exploitation ratio} (70\%) govern how each parent program is chosen, drawing either a random program from the island or an elite from the archive.
Each parent is presented to the LLM alongside 2 inspiration programs
sampled from the same island and the island's 3 highest-scoring programs.

\cat{Benchmark Suite}
Both Stage~1 and Stage~2 use the same benchmark suite, which includes TPC-C and AuctionMark applications, as well as three synthetic workloads targeting specific correctness scenarios: a \textit{banking} workload (concurrent fund transfers), a \textit{counter} workload (read-modify-write increments on a shared counter), and a \textit{deadlock} workload (cyclic resource-acquisition patterns).
In Stage~1, the \textit{Serializability Verifier} runs the benchmark suite within the generate-and-verify loop, checking only for conflict-serializability.
In Stage~2, the suite serves as the fitness function for evolutionary optimization, scoring each candidate on both correctness and throughput as described in Section~\ref{sec:design}.

\cat{TPC-C Application}
TPC-C~\cite{tpc-c} benchmark simulates a wholesale supplier executing five transaction types across multiple warehouses. Following prior CC evaluation practice~\cite{ren2016deadlockfree, guo2021bamboo, prasaad2020handling}, we focus on \textit{New Order} and \textit{Payment}, which together account for approximately \(88\%\) of the benchmark mix (\(45\%\) and \(43\%\), respectively) and are short update transactions that place the greatest stress on concurrency control.
We use a \(50\%\)/\(50\%\) split between the two transaction types, with \(1\%\) of transactions subject to random user aborts per the benchmark specification.
We evaluate TPC-C along two independent contention axes.
The first is a \textit{scalability benchmark}: we vary the warehouse count across six levels (\textit{very low} to \textit{extreme}), from 64 warehouses (low inter-transaction conflict) down to 1 warehouse (all transactions share the same key space).
The second is a \textit{hotspot benchmark}: we fix the warehouse count at 8 and vary the hotspot intensity across four levels, concentrating accesses onto a small fraction of hot records to stress intra-warehouse contention independently of scale.

\cat{AuctionMark Application}
AuctionMark benchmark from OLTP-Bench~\cite{DifallahPCC13} models an online auction platform with a mixed read-heavy and write-heavy transaction mix. Key contention points include high-bid records and item-status fields, which are accessed by multiple concurrent transaction types. We run AuctionMark at four contention levels by varying the Zipfian skew parameter applied to item and user access patterns, ranging from \textit{low} to \textit{extreme} hotspot concentration. A higher skew concentrates a larger fraction of transactions onto a small set of hot item and user records that directly increases read-write conflict at those rows.

\cat{Evaluation Environment}
The experiments are conducted on \textit{cascadelake} node of Chameleon Cloud~\cite{keahey2020chameleon}. This machine has 2 \textit{Intel Xeon Gold 6242 CPUs} at \textit{2.8 GHz} with \textit{96} virtual threads and is also equipped with 187 GB RAM and 10 Gigabit Ethernet.

We implemented a transactional engine in Java with an in-memory key-value store, in the style of DBx1000~\cite{yu2014ccEvaluation}, that uses the generated \ourSystem\ protocol as its concurrency control mechanism. Each candidate protocol is evaluated by a concurrent Java benchmark running 64 parallel worker threads.
Each scenario runs for 30 seconds to collect throughput and correctness traces.
Serializability is checked after each scenario using precedence-graph cycle detection.

\subsection{Stage~1 Results}

We ran Stage~1 on TPC-C and AuctionMark and collected the first five protocols produced for each application---SALVO, LP-OCC, PESC, DCOV, and SADL for TPC-C, and TREAD, WGC, HSL, PVL, and SAPL for AuctionMark (described in detail in Section~\ref{sec:protocols}). We evaluate two aspects of the ten generated protocols: how their designs align with known CC families and how structurally similar the protocols are to each other.

\begin{figure}[t]
    \centering
    \includegraphics[width=0.8\linewidth]{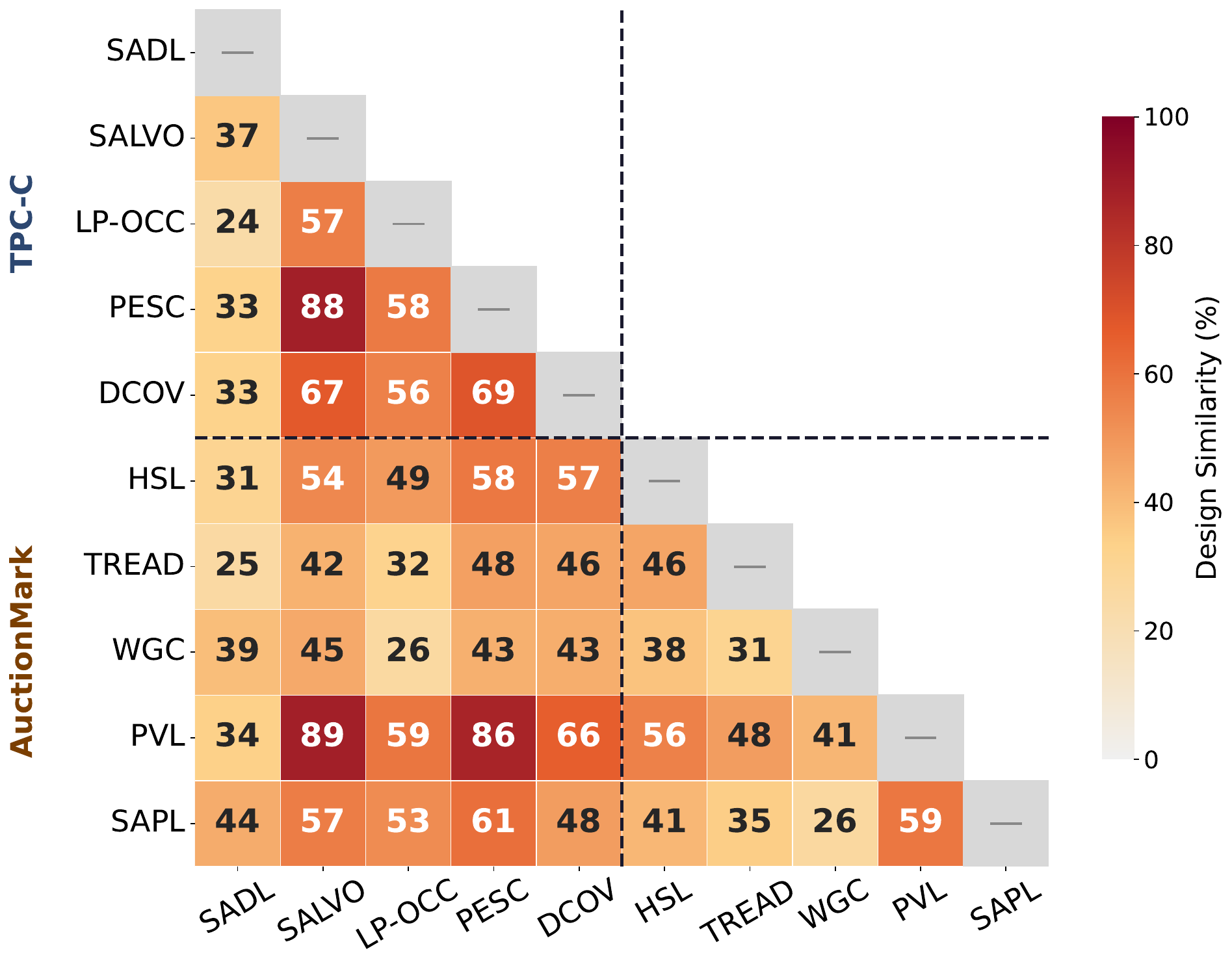}
    \caption{Pairwise design similarity between the ten generated CC protocols. 
\vspace{-1.5em}}
    \label{fig:similarity}
\end{figure}

\begin{figure*}[t]
    \centering
    \includegraphics[width=\textwidth]{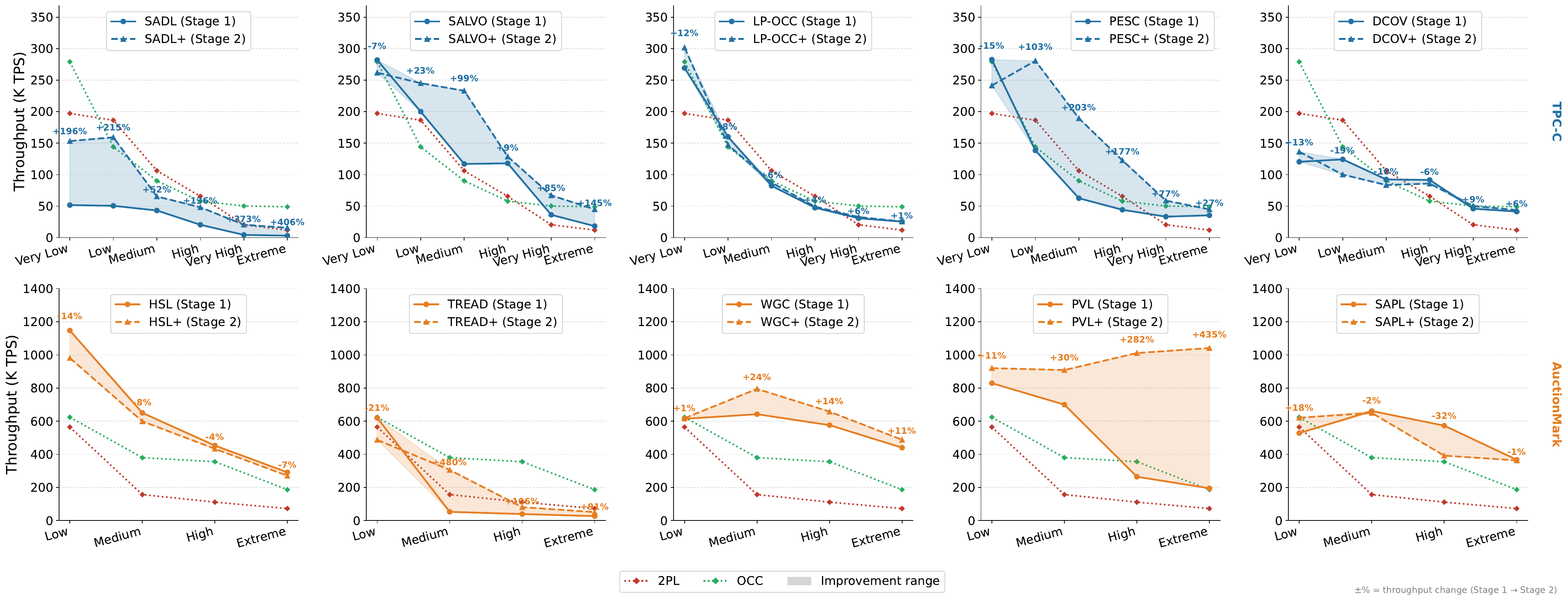}
    \caption{Throughput in thousands of transactions per second (K\,TPS) before (Stage~1) and after (Stage~2) evolutionary optimization across six contention levels for TPC-C protocols and four contention levels for AuctionMark protocols. Shaded area shows the improvement range; 2PL and OCC baselines shown for reference. Percentages denote the per-level throughput change from Stage~1 to Stage~2.}
    \label{fig:improvement}
\end{figure*}

\cat{Protocol Diversity}
To assess structural diversity, we describe each protocol as a binary feature vector, where each entry records whether the protocol uses a particular design technique (1) or not (0). We define 150 such features, grouped into 12 design dimensions---pessimistic locking~\cite{bernstein1987concurrency}, optimistic validation~\cite{kung1981occ}, conflict detection timing, multi-version storage~\cite{bernstein1983multiversion}, correctness guarantees, SSI-style anti-dependency tracking~\cite{cahill2008ssi}, timestamp ordering~\cite{bernstein1980timestamp}, interval-based commit timestamps~\cite{yu2016tictoc}, deterministic pre-ordering~\cite{thomson2012calvin, faleiro2014bohm}, partition-based execution~\cite{kallman2008hstore}, commutativity-aware conflict resolution~\cite{oneil1986escrow}, and implementation-level features.
Pairwise similarity between two protocols is computed using a Jaccard-style metric~\cite{jaccard1912flora} over positive criteria:
\begin{equation}
    \text{sim}(A, B) \;=\; \frac{|\{c : A_c = 1 \wedge B_c = 1\}|}{|\{c : A_c = 1 \vee B_c = 1\}|} \times 100
    \label{eq:jaccard}
\end{equation}
The score is the fraction of techniques both protocols adopt (numerator) among all techniques used by either protocol (denominator), scaled to a percentage. Shared absences (both 0) are excluded, so similarity reflects genuine structural alignment rather than coincidental non-adoption of features.

Figure~\ref{fig:comparison} shows the design profile of each generated protocol as a radar chart, where each axis represents the Jaccard similarity to a reference CC family (2PL~\cite{bernstein1987concurrency}, OCC~\cite{kung1981occ}, MVCC~\cite{bernstein1983multiversion}, SSI~\cite{cahill2008ssi}, Silo~\cite{tu2013speedy}, H-Store~\cite{kallman2008hstore}, Escrow~\cite{oneil1986escrow}, Calvin~\cite{thomson2012calvin}, TO~\cite{bernstein1980timestamp}, TicToc~\cite{yu2016tictoc}, Hekaton~\cite{larson2011high}).
No generated protocol collapses onto a single reference: all ten exhibit a distinct multi-axis profile, borrowing selectively from two to four reference families without reproducing any one of them.

Among TPC-C protocols (blue), SALVO, LP-OCC, PESC, and DCOV all cluster strongly around the OCC/Silo/TicToc family (47--55\% on those axes), consistent with optimistic validation designs tailored to short-update workloads.
SADL diverges from the rest of the TPC-C protocols, with the highest Escrow (48\%) and Calvin (45\%) similarity of any generated protocol, which reflects a commutative-operation and reservation-based design rather than a conventional optimistic approach.

Among AuctionMark protocols (orange), PVL is the most Silo-dominant (59\%), while TREAD diverges sharply from the rest with its highest scores on Hekaton (52\%) and SSI (48\%), indicating a version-based, anti-dependency-tracking design.
WGC stands out as the most H-Store-similar protocol (40\%), the only partition-style design in the cohort, while SAPL combines a notable Escrow affinity (46\%) with broad OCC and Silo characteristics.

\cat{Protocol Similarity}
Figure~\ref{fig:similarity} shows the pairwise Jaccard similarity between all ten generated protocols. This heatmap reveals a tight structural cluster: SALVO (TPC-C), PESC (TPC-C), and PVL (AuctionMark) are nearly identical ($\ge$86\% pairwise), with DCOV and LP-OCC also closely related (56--69\% to the cluster core). This cluster spans both target workloads, showing that the Proposer can independently converge on similar OCC/Silo-family designs when both workloads reward optimistic execution.

Three protocols stand clearly apart from the rest, with SADL reaching at most 44\% similarity to any peer, TREAD 48\%, and WGC 45\%---each representing a design family (Escrow/Calvin, Hekaton/SSI, and H-Store partition, respectively) that no other generated protocol closely replicates. This combination of a tight optimistic cluster and three distinctly separated protocols confirms that Stage~1 produces genuine structural variety rather than uniformly sampling a single region of the protocol space, a spread the pipeline actively encourages by prompting against redundancy (Section~\ref{sec:design}).

\begin{figure*}[t]
    \centering
    \includegraphics[width=\textwidth]{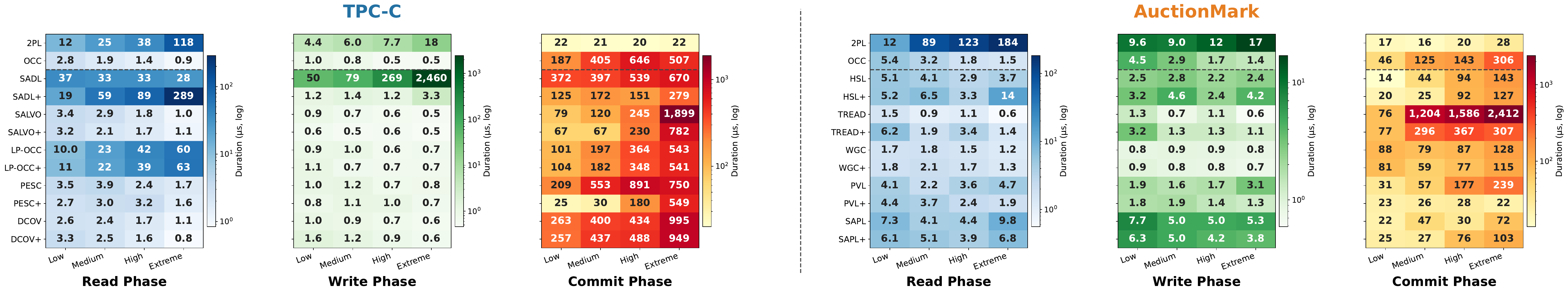}
    \caption{Mean read, write, and commit phase duration ($\mu$s) for all protocols before and after Stage~2 optimization, at four hotspot intensity levels. A \texttt{+} suffix on a protocol name denotes its post-Stage~2 variant.\vspace{-0.5em}}
    \label{fig:phase}
\end{figure*}

\subsection{Stage~2 Results}
\label{sec:stage2}

Stage~2 evolutionary optimization improves throughput for the majority of generated protocols across both workloads.
Figure~\ref{fig:improvement} shows per-protocol throughput curves before and after optimization for both workloads, with throughput (K\,TPS) on the $y$-axis and contention level on the $x$-axis; each protocol is plotted as a Stage~1 and a Stage~2 curve, the shaded band between them marks the improvement, and the 2PL and OCC baselines are overlaid for reference. Figure~\ref{fig:phase} decomposes execution time into read, write, and commit phases to expose the source of each improvement, reporting the mean per-phase duration (in $\mu$s) for every protocol before and after Stage~2 across four hotspot intensity levels.

\subsubsection{TPC-C Protocols}

\cat{Throughput Improvement}
Among TPC-C protocols, Stage~2 produces the largest gains for SADL and PESC, moderate contention-dependent gains for SALVO, and marginal changes for LP-OCC and DCOV (Figure~\ref{fig:improvement}).
SADL improves by $52--406\%$ across all six contention levels, the largest absolute gains in the TPC-C cohort.
SALVO shows a modest regression at very low contention ($-7\%$) but strong gains at high and extreme contention ($+85\%$ and $+145\%$), consistent with optimizations that pay off only when the commutative-write fast path is heavily exercised.
PESC shows a $-15\%$ regression at very low contention but recovers to $+77$--$203\%$ at higher levels, reflecting commit-path optimization that carries overhead at low concurrency.
LP-OCC and DCOV show near-flat or slightly negative changes ($-19\%$ to $+13\%$), indicating that Stage~1 already found a near-optimal implementation for these OCC-family designs.
All five TPC-C protocols outperform both the 2PL and OCC baselines at medium and higher contention levels after Stage~2.

\cat{Phase-Level Analysis}
Figure~\ref{fig:phase} reveals that the large SADL improvement originates from a critical write-phase defect: the Stage~1 SADL implementation exhibits a write phase of 50--2{,}460\,$\mu$s across contention levels, growing catastrophically to over 2\,ms at extreme hotspot contention.
Stage~2 identifies and eliminates this write-path bottleneck, reducing the write phase to 1.2--3.3\,$\mu$s, a $3$ orders-of-magnitude improvement at extreme contention.
PESC undergoes a similarly dramatic commit-phase transformation: the Stage~1 commit overhead of 209--891\,$\mu$s is reduced to 25--549\,$\mu$s after Stage~2, with the low-contention case dropping by $8\times$, reflecting the elimination of redundant validation work on the conflict-free fast path.
In contrast, LP-OCC and DCOV phase times are largely unchanged between Stage~1 and Stage~2, confirming that the small throughput deltas reflect implementation maturity rather than missed optimization opportunities.
The OCC baseline exhibits a characteristic validation bottleneck in the commit phase (187--646\,$\mu$s), while 2PL shows an escalating read phase (12--118\,$\mu$s) under increasing contention due to lock-wait latency.

\subsubsection{AuctionMark Protocols}

\cat{Throughput Improvement}
The AuctionMark cohort exhibits a wider range of Stage~2 outcomes than TPC-C (Figure~\ref{fig:improvement}).
TREAD undergoes the most dramatic recovery: a $-21\%$ regression at low contention is accompanied by $+480\%$, $+106\%$, and $+91\%$ gains at medium, high, and extreme contention, transforming a protocol that was barely functional at moderate contention into the highest-throughput design in the AuctionMark cohort at those levels.
PVL improves monotonically with contention---$+11\%$ at low, $+282\%$ at high, and $+435\%$ at extreme---becoming the best-performing AuctionMark protocol at high and extreme contention after optimization.
WGC shows consistent modest improvements of $+1$--$24\%$ across all levels, reflecting incremental refinements without any single dominant fix.
HSL degrades slightly at all four levels ($-4\%$ to $-14\%$), the only AuctionMark protocol where Stage~2 produces a net negative outcome.
SAPL gains $+18\%$ at low contention but regresses by $32\%$ at high contention, indicating that Stage~2 mutations that benefited the low-contention case introduced a new bottleneck under high concurrency.

\cat{Phase-Level Analysis}
Figure~\ref{fig:phase} identifies the commit phase as the dominant bottleneck in the AuctionMark cohort, mirroring the TPC-C findings.
TREAD's Stage~1 commit phase grows from 76\,$\mu$s at low contention to 2{,}412\,$\mu$s at extreme---a super-linear growth in commit overhead under increasing contention.
Stage~2 reduces this to a flat 77--367\,$\mu$s profile, eliminating the super-linear scaling behavior and enabling the large throughput recovery observed in Figure~\ref{fig:improvement}.
PVL shows a complementary pattern: its Stage~1 commit phase grows to 239\,$\mu$s at extreme contention, while Stage~2 achieves near-constant commit overhead of 22--28\,$\mu$s across all levels---comparable to the 2PL baseline---explaining the $+435\%$ throughput gain at extreme contention.
HSL, WGC, and SAPL exhibit commit phases in the 14--143\,$\mu$s range even at Stage~1, leaving little room for commit-path optimization; the modest throughput changes for these protocols are consistent with this observation.
Read and write phase times remain below 10\,$\mu$s for most AuctionMark protocols across all contention levels, confirming that the bottleneck is exclusively in the commit phase rather than in data access or buffer management.

\vspace{-5pt}
\section{Generated Concurrency Control Protocols}
\label{sec:protocols}

In this section, we describe the ten concurrency control protocols generated by \ourSystem: five targeting TPC-C and five targeting AuctionMark. Each protocol is described below with its design overview, read/write/commit strategy, and the improvements found by Stage~2. The correctness argument stated for each protocol is the conflict-serializability justification \ourSystem\ produced during design and confirmed in Stage~1 verification.

\vspace{-5pt}
\subsection{TPC-C Specific Protocols}
\label{sec:tpcc-protocols}

\subsubsection{SADL -- Semantic-Aware Deterministic Locking}
\label{sec:sadl}

SADL\footnote{All protocol names are generated by the \ourSystem\ Proposer agent (an LLM), not chosen by the authors.} partitions the TPC-C key space by warehouse ID. Each partition is managed by a single-threaded scheduler that assigns each transaction a monotonic \emph{intra-partition sequence number} (IPSN), establishing a pre-defined serial order. Concurrency is maximized through three lock modes: shared (S) for reads, exclusive (X) for non-commutative writes, and an \emph{ordered-commutative} (OC) lock for arithmetic writes. OC-locks are mutually compatible, allowing concurrent commutative updates to the same field while the commit queue enforces the IPSN order.

SADL guarantees conflict-serializability, as the pre-assigned IPSN fixes a per-partition serial order and applying buffered writes in that order makes the committed state equivalent to a serial execution. Its performance gains rely on a largely partition-friendly workload; cross-partition transactions remain correct but forgo the benefit of independent per-partition scheduling.

\cat{Read/Write/Commit}
Reads acquire an S-lock and return the store value. Writes to order-ID counter fields are non-commutative and acquire an X-lock; all other writes (year-to-date balance updates, stock quantity adjustments) are commutative arithmetic operations and acquire an OC-lock. All writes are buffered locally until commit. Commit places the transaction in a per-partition priority queue; the scheduler drains this queue in strict IPSN order, applies buffered writes, and releases locks. Deadlocks are prevented because a transaction may wait for a lock held by another only when its IPSN is larger, producing an acyclic waits-for graph.

\cat{Stage~2 Improvements}
The optimizer fixed an IPSN assignment race that allowed duplicate sequence numbers, an initialization error that caused partitions to share a single scheduler, and a missing wake-up of waiting transactions in the per-partition commit scheduler that produced head-of-line blocking under burst loads. All three are low-level concurrency defects that suppressed throughput rather than algorithmic changes; removing them unlocked the per-partition and OC-lock parallelism the design already intended, accounting for SADL's large throughput gain.

\subsubsection{SALVO -- Semantic-Aware Latching and Optimistic Delta Application}
\label{sec:salvo}

SALVO is a hybrid OCC protocol that dispatches each read and write to one of three specialized modes selected by runtime key-pattern matching. District-level order counter fields are served by lock-free atomic increments. Customer balance reads that are unconditionally overwritten by the same transaction are excluded from validation (blind-read elimination). All remaining fields follow standard OCC versioned reads. Three write modes mirror this structure: new-order records are private inserts with version~0, year-to-date and balance fields accumulate commutative deltas, and all other writes are buffered as versioned values.

SALVO provides conflict-serializability, combining OCC validation on the optimistic path with order-independent commutative deltas and sorted-order commit locking for deadlock freedom.

\cat{Read/Write/Commit}
A read returns the requested value and records its version for later validation, except on the two fast paths, where counter reads use an atomic fetch-and-add and blind balance reads skip version tracking. A write is buffered locally in its mode---a versioned value, a commutative delta, or a private insert---and applied only at commit. Commit locks all written and delta keys in sorted order for deadlock freedom, validates that no read version has changed, and then installs the versioned writes, applies the deltas atomically, and adds the private inserts before releasing the locks.

\cat{Stage~2 Improvements}
The optimizer widened the commutative delta set from warehouse-level to district-level year-to-date and order-counter fields (the actual hot spots at low warehouse counts) and corrected a double-apply bug that re-applied deltas during checkpoint replay.

\subsubsection{LP-OCC -- Lock-Protected Optimistic Concurrency Control}
\label{sec:lp-occ}

LP-OCC is a classic OCC implementation that eliminates reader-writer races on value-version pairs by briefly locking each key during the read phase. Each record has a dedicated per-key lock in a global lock manager. Transactions execute optimistically, buffering all writes locally; the shared store is only modified at commit.

LP-OCC provides conflict-serializability, since the read-phase lock makes each read observe a fully installed value and version, and holding locks over the whole read and write set at commit makes validation and installation atomic.

\cat{Read/Write/Commit}
A read briefly locks the key, captures the value and its version together, and releases the lock, so the transaction always sees a consistent snapshot; repeat reads are served from a local cache. A write is buffered locally, and if the key was not already read, a consistent read is issued first so that every written key carries a version dependency and no update is lost. Commit locks the union of the read and write set in sorted order (keeping it deadlock-free), validates that no recorded version has changed, and, if validation passes, installs the buffered writes and bumps their versions before releasing the locks.

\cat{Stage~2 Improvements}
The optimizer stripped avoidable overhead from the commit path, removing synchronization from per-transaction data structures and redundant allocation when ordering keys for locking. It also routed each write to the store's matching operation---insert, numeric increment/decrement, or overwrite.

\subsubsection{PESC -- Phased Execution and Semantic Commit}
\label{sec:pesc}

PESC is a hybrid concurrency control that routes each write to one of two paths based on key semantics. Year-to-date and balance fields are commutative and accumulate signed deltas; all other fields are non-commutative and buffer absolute values. A per-transaction flag tracks whether any non-commutative write has occurred, enabling a blind-read optimization that skips version tracking for customer balance reads in New Order transactions (which unconditionally overwrite those fields).

PESC is conflict-serializable, as the non-commutative path validates reads and then holds write locks through installation so no concurrent commit can interfere, while commutative delta writes are order-independent.

\cat{Read/Write/Commit}
Reads check a local snapshot cache; misses fetch the value and version stamp from the store. Commutative writes compute a delta against the snapshot and merge it into a delta set; non-commutative writes buffer the intended value in a write set. Commit acquires per-key locks in sorted order, validates the read set, re-reads current locked values for stock fields and recomputes their deltas relative to the snapshot (absorbing concurrent writes that committed between the read and the lock), installs all writes, and releases locks.

\cat{Stage~2 Improvements}
The optimizer removed a bypass that routed order-ID counter writes directly through an atomic variable outside the transaction protocol (leaving those writes untracked in version state), narrowed the locking scope to write keys only to reduce contention, and routed every write through the store's typed insert, update, and increment operations.

\subsubsection{DCOV -- Dual-Counter Optimistic Validation}
\label{sec:dcov}

DCOV is a semantic OCC protocol that replaces the conventional single version number per record with two monotonic counters: an \emph{Exclusive-Modification Counter} (EMC) incremented by non-commutative writes and a \emph{Commutative-Modification Counter} (CMC) incremented by delta writes. Each record is stored as a (value, EMC, CMC) triple. A per-warehouse commit lock serializes all commits within a partition; validation confirms both counters match for every non-blind read key.

DCOV is conflict-serializable, since the per-warehouse commit lock makes each transaction's validation and installation atomic, and matching both counters confirms that no committer has modified its non-blind reads since it observed them.

\cat{Read/Write/Commit}
Reads check the local write set, then a per-transaction cache, then the store. Writes to year-to-date and balance fields compute a signed delta; all other writes buffer an absolute value and mark the transaction as having an exclusive write. A read is classified as \emph{blind} if the transaction has already performed at least one exclusive write and the key is a customer balance field that will be unconditionally overwritten; blind reads are cached but excluded from counter validation. Commit acquires the per-warehouse commit lock, validates all non-blind read-set entries against both EMC and CMC, installs each write (incrementing EMC for exclusive writes, CMC for delta writes), and releases the lock.

\cat{Stage~2 Improvements}
The optimizer removed synchronization from per-transaction data structures that are never shared across threads, added a validation-result cache to reuse fetched records across the validation and write phases, and fixed partition-key parsing to correctly route global item records.

\subsection{AuctionMark Specific Protocols}
\label{sec:auction-protocols}

\subsubsection{HSL -- Hierarchical Speculative Locking}
\label{sec:hsl}

HSL groups records into coarse-grained \emph{logical entities} and enforces concurrency at entity granularity. Transactions execute speculatively in parallel without acquiring any lock during execution; they register read and write intents on entity-level data structures. Concurrency control is enforced entirely at commit through firm-lock acquisition and proactive invalidation of all concurrent speculators on written entities.
HSL guarantees conflict-serializability, as a committing writer holds firm locks on every entity it touched and proactively aborts any transaction that speculatively read a written entity, so no transaction can commit on a stale read.

\cat{Read/Write/Commit}
Reads register the transaction in the entity's speculative-reader map and fetch the value directly from the store without acquiring any lock. Writes register in the speculative-writer map and buffer the key-value pair locally. Read-only transactions (empty write set) skip locking entirely and commit immediately after deregistration. Writing transactions sort their read and written entity IDs lexicographically and acquire each entity's firm lock in that order. Writes are applied atomically; every transaction that read from a written entity is then proactively aborted before locks are released.

\cat{Stage~2 Improvements}
The optimizer expanded entity-prefix coverage to the full AuctionMark schema (adding item-bid, item-purchase, user-item, and composite counter entity types), reducing false conflicts from unrecognized keys falling through to the single global lock entity, and fixed cleanup to remove transactions from the correct speculative map rather than cross-probing both maps for every entity.

\subsubsection{TREAD -- Transactional Read Epoch-based Abort \& Deltas}
\label{sec:tread}

TREAD is a multi-version protocol built on an MVCC backbone, where each key holds a version chain and reads are non-blocking, serving the latest version committed before the transaction's start timestamp. Item-status reads additionally register the transaction in the read version's dependency set, while high-contention counter fields---bid-count, rating, and per-item-user counters---bypass the version chain and accumulate as commutative signed deltas. These mechanisms lift MVCC's snapshot isolation toward serializability, since a transaction that writes a status field aborts the dependents registered on the overwritten version before installing its own, breaking the anti-dependency cycles behind write-skew, while commutative deltas commit in any order.

\cat{Read/Write/Commit}
Reads walk the version chain for the latest visible version; item-status reads additionally register in the version's dependency set. Commutative writes append a signed delta to a local buffer; all others buffer absolute values. At commit, each standard write acquires a hash-stripe latch---rather than a lock per record, keys are hashed onto a fixed-size array of latches, so only keys colliding on the same entry contend---then clears and processes the current version head's dependency set (proactively aborting stale readers), installs a new version node, and immediately releases the latch. Deltas are applied atomically without acquiring any latch, so concurrent delta commits on the same counter field never block each other.

\cat{Stage~2 Improvements}
The optimizer narrowed dependency invalidation from a full version-chain scan to only the current version head (reducing cost from $O(V)$ to $O(1)$ per key, where $V$ is the key's version-chain length), expanded the latch array from 128 to 1024 entries to reduce collision probability, and extended the commutative field set to cover order-ID counters, quantity, total, balance, and year-to-date fields.

\subsubsection{WGC -- Workflow-Aware Graph Coloring}
\label{sec:wgc}

WGC transforms concurrency control from runtime conflict detection into pre-execution scheduling. Transactions are classified into three semantic colors: GREEN (read-only), YELLOW (writes that touch no status field), or RED (writes that change an item's status). A pool of 128 single-threaded \emph{Entity Execution Coordinators} (EECs) partitions the entity space by hash. Each EEC enforces a color-compatibility matrix: multiple GREENs overlap freely; a single YELLOW may overlap with concurrent GREENs; REDs require the entity to be fully idle. Incompatible transactions are held in a hold-back queue and re-evaluated after each completion.
WGC is conflict-serializable, since transactions that conflict on an entity are dispatched to the same single-threaded EEC and run one at a time, while the color rules let only non-conflicting transactions overlap.

\cat{Read/Write/Commit}
Reads are client-side and non-blocking. Bid-count, rating, and per-item-user counter writes compute and buffer a signed delta; all other writes buffer an absolute value. At commit, the transaction's color is determined (GREEN if no writes exist, RED if any write touches an item-status field, YELLOW otherwise), it is submitted to its EEC, and the caller blocks until the EEC applies the writes and signals completion.

\cat{Stage~2 Improvements}
The optimizer replaced an $O(n)$ scan over active transactions in the compatibility check with three running per-color counts, and restructured the hold-back queue so that waiting transactions are re-evaluated in FIFO order after each completion.

\subsubsection{PVL -- Optimized Phased Validation Locking}
\label{sec:pvl}

PVL combines lock-free reads with a three-path commit strategy differentiated by write semantics. Each key carries a version counter. Reads record versioned snapshots; writes are classified as commutative deltas or standard absolute values, routing the commit to the lightest correct path for the transaction's write footprint.
PVL is conflict-serializable, since every commit path validates that no version in the read snapshot has changed, and the standard path additionally holds write locks across validation and installation so no concurrent writer can intervene.

\cat{Read/Write/Commit}
Reads sample the version counter and value, cache the result as a versioned snapshot, and return without acquiring any lock. Bid-count, rating, and per-item-user counter writes accumulate a signed delta; all other writes buffer absolute values. Commit takes one of three paths: (1)~\emph{read-only} validates version snapshots with no locks acquired; (2)~\emph{delta-only} validates versions and applies deltas atomically without acquiring any write lock, making high-frequency counter transactions nearly contention-free; (3)~\emph{standard} acquires write locks in sorted key order, validates, installs absolute writes, and applies deltas.

\cat{Stage~2 Improvements}
The optimizer fixed a type-handling bug in the read path that failed on certain numeric encodings and caused spurious aborts, normalizing all integer-valued fields to a common type; this repair simplified the write-key union computed at commit.

\subsubsection{SAPL -- State-Aware Phased Locking}
\label{sec:sapl}

SAPL combines pessimistic entity latching with OCC version validation in two layers. Transactions acquire shared or exclusive read/write latches on logical item entities throughout their lifetime, serializing state-transitioning operations. A version-validation layer at commit serves as a correctness backstop for remaining conflicts. Commutative counter updates are buffered as signed deltas and applied atomically at commit.
SAPL is conflict-serializable through its two layers, with entity latches serializing state transitions on an item and commit-time version validation under sorted key locks aborting any transaction whose reads changed before its writes install.

\cat{Read/Write/Commit}
Reads to item-centric entity fields acquire a shared read latch with a non-blocking attempt that aborts on failure, and sample the version counter around the store fetch to detect torn reads. Bid-count, rating, and comment-counter writes accumulate a signed delta, while balance and year-to-date writes take the standard OCC path; item-status writes escalate the entity latch from shared to exclusive (with a version check between release and re-acquire to detect intervening modifications). Commit acquires per-key locks in sorted order, validates all read dependencies, installs absolute writes and deltas, and releases all locks and latches.

\cat{Stage~2 Improvements}
The optimizer moved balance and year-to-date fields off the commutative delta path and back onto standard OCC validation, since treating their updates as blind commutative deltas violated per-record invariants; it also narrowed pessimistic latching to item records only and simplified how keys are mapped to entities.

\vspace{-5pt}
\section{Related Work}
\label{sec:Related_work}

\ourSystem\ builds on decades of theoretical and practical research on concurrency control protocols. There has been extensive research on distinct families of concurrency control protocols, each tailored to a specific hardware paradigm or workload. 

Two-Phase Locking (2PL)~\cite{bernstein1987concurrency} enforces conflict ordering through lock acquisition; Optimistic Concurrency Control (OCC)~\cite{kung1981occ} defers validation to commit time; Multi-Version Concurrency Control (MVCC)~\cite{bernstein1983multiversion} maintains version histories for non-blocking reads; and Serializable Snapshot Isolation (SSI)~\cite{cahill2008ssi} extends snapshot isolation to full serializability via anti-dependency tracking. Modern in-memory systems such as Silo~\cite{tu2013speedy}, Hekaton~\cite{larson2011high}, TicToc~\cite{yu2016tictoc}, Cicada~\cite{lim2017cicada}, and Bamboo~\cite{guo2021bamboo} optimize these foundations for multi-core hardware, while deterministic protocols such as Calvin~\cite{thomson2012calvin} and BOHM~\cite{faleiro2014bohm} achieve predictable performance through pre-ordered execution. Brook-2PL~\cite{habibi2026brook} targets high-contention hotspot workloads by statically analyzing a graph-based dependency structure (SLW-Graph) to enable deadlock-free lock acquisition and partial transaction chopping for early lock release. IC3~\cite{wang2016ics3} similarly exploits static transaction analysis to reduce conflict footprint through fine-grained chopping. Each of these families excels under specific workload conditions but none adapts to a new workload without manual re-engineering.

\cat{Hybrid CC Protocols}
Prior work has explored combining multiple concurrency control mechanisms to improve performance across varying workload conditions. Static and Dynamic OCC~\cite{yu1992analysis} switch from OCC to locking after the first retry. MOCC~\cite{wang2016mocc} and Hsync~\cite{shang2016hsync} propose hybrid 2PL--OCC protocols. ACC~\cite{tang2017acc} goes further by adaptively selecting from a broader range of protocols at runtime. Tebaldi~\cite{su2017tebladi} groups transactions and assigns each group a different CC protocol. CormCC~\cite{tang2018CormCC} proposes a more general framework by formalizing CC into four phases, allowing each operation to adopt any protocol’s policy as long as all protocols execute each phase in the same order.

These approaches, however, are generally scoped to combinations of OCC and 2PL and require programmers to supply heuristics for choosing execution policies per operation. They are designed for dynamic workload adaptation rather than application-specific protocol design, and they do not exploit application-level transaction semantics. \ourSystem\ differs by generating a complete, application-specific protocol implementation.

\cat{Learned CC Protocols}
The observation that a single protocol cannot satisfy all contention levels and workload patterns led to the development of learned concurrency control protocols.

Polyjuice~\cite{wang2021polyjuice} models concurrency control as a dynamic policy table, defining the execution state as a high-dimensional combination of transaction type and data access patterns. The system searches a policy space of fine-grained actions to determine whether a transaction should wait, expose a speculative write, or abort. However, it does not support general transactions and treats all transactions accessing the same data uniformly, which can produce suboptimal decisions with high abort rates.

NeurCC~\cite{pan2026modeling} advances this direction by modeling the CC mechanism as a continuous learnable function that maps dynamic database states directly to conflict-handling actions, coupled with Bayesian optimization and a graph reduction search algorithm. Both systems, however, generalize to a fixed action space across all workload types and do not account for application-specific semantics.

\ourSystem\ takes a fundamentally different approach. Rather than learning a policy over a predefined action space, it uses LLM agents to generate application-specific protocol implementations from scratch, producing protocols whose structure and semantics are tailored to the target application.

\cat{LLM-driven Agents}
Large language models trained on code can generate executable programs from natural-language specifications~\cite{chen2021codex}, and recent work embeds them in agentic loops that reason, act, and repair their output under feedback~\cite{yao2023react, shinn2023reflexion, madaan2023selfrefine}. SWE-agent~\cite{yang2024sweagent} edits real codebases for automated software engineering, while multi-agent frameworks such as MetaGPT~\cite{hong2024metagpt} and AutoGen~\cite{wu2023autogen} assign specialized roles to collaborating agents. These systems target general programming tasks and validate their output against unit tests, which cannot certify the correctness of a concurrency control protocol. \ourSystem\ instantiates this agentic paradigm for CC protocol design, pairing specialized proposal and generation agents with a serializability verifier that enforces a formal correctness criterion in the loop.

\cat{Evolutionary Optimization}
Evolutionary programming is a classical approach to program synthesis that uses mutation and crossover operators to evolve a population of candidate programs~\cite{langdon2002foundations, koza1994genetic}. Evolutionary algorithms have been applied to regression~\cite{ma2022evolving, schmidt2009distilling}, automated algorithm and scientific discovery~\cite{cranmer2023interpretable, chen2023symbolic}, and scheduling~\cite{zhang2021genetic}. However, these methods rely on handcrafted evolution operators that are difficult to design and may fail to capture domain-specific properties.

Recent work has improved on this by replacing traditional operators with LLMs~\cite{romera2024funSearch, novikov2025alphaevolve, openevolve, assumpccao2025codeevolve}. FunSearch~\cite{romera2024funSearch} pairs a pretrained LLM with a systematic evaluator in an evolutionary loop, allowing the LLM to propose program mutations guided by feedback. AlphaEvolve~\cite{novikov2025alphaevolve} extends this by enabling evolution over entire codebases, supporting multi-objective optimization, and using frontier LLMs as mutation engines. \ourSystem\ builds on OpenEvolve~\cite{openevolve}, an open-source implementation of AlphaEvolve, to drive Stage~2 optimization of CC protocol implementations.

Evolutionary optimization has recently been applied to systems research problems more broadly~\cite{cheng2025let, tagliabue2025aiForDistystem, cheng2026aiDatabases, erol2026makingDatabaseFaster}. Authors in~\cite{cheng2025barbarians} present case studies across cloud scheduling, load balancing, SQL query optimization, and transaction scheduling. They argue that the key challenge for AI in systems is problem formulation and strategic guidance rather than low-level implementation. \ourSystem\ extends this direction to CC protocol design, applying LLM-driven evolutionary optimization with correctness and throughput as joint objectives.

\section{Conclusion}
\label{sec:conclusion}

We presented \ourSystem, a two-stage LLM-driven pipeline that automatically designs, verifies, and optimizes application-specific concurrency control protocols with minimal human intervention.
Running \ourSystem\ on TPC-C and AuctionMark produced ten protocols, all of which are verified for conflict-serializability after Stage~1, with Stage~2 improving the combined correctness-throughput score for nine out of ten protocols, yielding average throughput gains of +50.6\% for TPC-C and +92.2\% for AuctionMark. 
The ten generated protocols span multiple design families, including pessimistic, optimistic, versioned, deterministic, and semantic designs, confirming that application-conditioned generation produces structural diversity rather than converging on a single template.
\ourSystem\ establishes correctness empirically rather than by formal proof, and integrating formal verification while extending the pipeline to distributed protocols is a natural next step.
\ourSystem\ proves that LLM-driven agentic pipelines can serve as practical tools for CC protocol design, producing correct and competitive protocols across diverse OLTP workloads.

\bibliographystyle{ACM-Reference-Format}
\bibliography{sample-base}

\end{document}